\documentclass[aps,pra,twocolumn,showpacs,amssymb]{revtex4}
\usepackage{amsmath2000}
\usepackage{graphicx}
\usepackage{amsfonts}

\def\be{\begin{equation}}
\def\ee{\end{equation}}
\def\bea{\begin{eqnarray}}
\def\eea{\end{eqnarray}}
\def\M{{\cal M}}
\def\N{{\cal N}}

\def\O{{\cal O}}
\def\E{{\cal E}}
\def\H{{\cal H}}
\def\K{{\cal K}}
\def\T{{\cal T}}
\def\tr{\mbox{ tr}}
\def\bra#1{\langle#1|} \def\ket#1{|#1\rangle}
\def\braket#1#2{\langle#1|#2\rangle}

\def\proj#1{\ket{#1}\!\bra{#1}}

\def\ra{\rangle}

\newcommand{\mattwo}[4]{\left(
        \begin{array}{rr}{#1}&{#2}\\{#3}&{#4}\end{array}\right)}

\begin{document}

\title{Non--local Hamiltonian simulation \\ assisted by local
operations and classical communication}

\author{G. Vidal$^{1,2}$ and J. I. Cirac$^{1}$}

\affiliation
{$^1$Institut f\"ur Theoretische Physik, Universit\"at Innsbruck,
A-6020 Innsbruck, Austria\\
$^2$Institute for Quantum Information, California Institute of Technology, Pasadena, CA 91125, USA}

\date{\today}

\begin{abstract}

Consider a set of $N$ systems and an arbitrary interaction
Hamiltonian $H$ that couples them. We investigate the use of
local operations and classical communication (LOCC), together
with the Hamiltonian $H$, to simulate a unitary evolution of the
$N$ systems according to some other Hamiltonian $H'$. First, we
show that the most general simulation using $H$ and LOCC can be
also achieved, with the same time efficiency, by just interspersing the evolution of $H$ with
local unitary manipulations of each system and a corresponding
local ancilla (in a so-called LU+anc protocol). Thus, 
the ability to make local measurements and to communicate classical information does not help in non--local Hamiltonian simulation. 
Second, we show that both for the case of two $d$-level systems ($d>2$), or for that of a setting with more than two systems ($N>2$), LU+anc protocols are more
powerful than LU protocols. Therefore local ancillas are a useful resource for non--local Hamiltonian simulation. Third, we use results of majorization theory to explicitly solve the problem of optimal simulation of two-qubit
Hamiltonians using LU (equivalently, LU+anc, LO or LOCC).

\end{abstract}


\maketitle

\section{Introduction}

The problem of using a given non--local Hamiltonian $H$ and some
class of local operations to simulate another non--local Hamiltonian
$H'$ has very
recently attracted the attention of several authors in quantum information science \cite{Dur,Dod,Woc,wir,Woc2,Nie,Woc3}. Nonetheless, average Hamiltonian techniques, a basic ingredient in non--local Hamiltonian simulation, have been studied for many years in control theory \cite{control}, and are commonly used in the area of nuclear magnetic resonance \cite{NMR}.

From the perspective of quantum information science, non--local Hamiltonian simulation sets a frame for the parameterization of the non--local resources
contained in multi--particle Hamiltonians, very much in the line
of thought pursued to quantify the entanglement of multi--particle quantum
states. In the most common setting, fast local unitary operations LU are performed on a series of systems to effectively modify the Hamiltonian $H$ that couples them. A remarkable result is the qualitative equivalence of all bipartite interactions under LU \cite{Dod,wir,Woc2,Nie,Woc3}. This can be shown to imply that any Hamiltonian $H$ with pairwise interactions between some of the systems can simulate any other Hamiltonian $H'$ consisting of arbitrary pairwise interactions between the same systems.

At a quantitative level, the time--efficiency with which a Hamiltonian $H$ 
is able to simulate a Hamiltonian $H'$ can be used as a criterion to endow
the set of non--local Hamiltonians with a (pseudo) partial order structure,
that allows to compare the non--local capabilities of $H$ and
$H'$ \cite{wir}. For two-qubit Hamiltonians, simulations using LU or arbitrary local operations LO have been shown to yield the same optimal time efficiencies, and the resulting partial order structure has been computed
explicitly. This has led to the necessary and sufficient conditions
for $H$ to be able to simulate $H'$ {\em efficiently} for
infinitesimal times, that is, the conditions under which the use
of $H$ for time $t$ allows to simulate $H'$ for the same time
$t$, in the small time $t$ limit. Equivalently, this result shows how to
{\em time--optimally} simulate $H'$ with $H$, in the sense of
achieving the maximal simulation ratio $t'/t$, where $t$ is the time of interaction
$H$ that it takes to simulate interaction $H'$ for a time $t'$.

\subsection{Ancillary systems, generalized local measurements and classical communication in non--local Hamiltonian simulation}

The aim of this paper is to elucidate the role a number of resources play in the simulation of non--local Hamiltonians. Relatedly, we seek at establishing equivalences between different classes of operations that may be used in a simulation protocol. 

We first address the question whether classical communication (CC) between the systems is useful in non--local Hamiltonian simulation. 
Recall that in protocols that include local measurements, the ability to communicate which outcome has been obtained in measuring one of the systems allows for subsequent operations on other systems to depend on this information. Now, can this ability be used in non--local Hamiltonian simulation to enlarge the set of achievable simulations? Suggestively enough, the answer is yes in the closely related problem of converting one non--local gate into another non--local gate using LO. For instance, a series of two--qubit gates $U$ exist such that they can be achieved by performing a C--NOT gate and LOCC but can not be achieved by a C--NOT and LO \cite{gates}.

We also study the advantage of using ancillary systems in simulation protocols, as well as performing general local operations instead of just local unitary transformations. Altogether, our analysis refer to the following classes of transformations: 

\begin{itemize}
\item LU = local unitary operations, 
\item LU+anc = local unitary operations with ancillas, 
\item LO = local operations \cite{note1}, 
\item LOCC = local operations with classical communication.
\end{itemize}

\subsection{Results}

This paper contains the following three main results concerning
the simulation of non--local Hamiltonian evolutions for infinitesimal
times:

($i$) LOCC (or LO) simulation protocols can be reduced to LU+anc
simulation protocols. That is, for $N$-particle Hamiltonian
interactions $H$ and $H'$, any protocol that simulates $H'$
using $H$ and LOCC (or LO) can be replaced, without changing its time
efficiency, with a protocol involving only $H$ and local unitary
transformations. Each local unitary transformation may be performed jointly 
on one of the $N$ systems and a local ancilla.

($ii$) Apart from exceptional cases such as that
of two-qubit Hamiltonians \cite{wir} ---in which any LU+anc
protocol can be further replaced with an even simpler protocol that uses only LU on each
qubit---, the use of ancillas is, in general, advantageous. This
is proven by constructing explicit examples of LU+anc protocols 
where ancillas are used to obtain simulations that
cannot be achieved with only LU operations, both in the case of two
$d$-level systems ($d>2$) and in the case of $N>2$ systems.

($iii$) For two-qubit Hamiltonians, we use results of
majorization theory to recover the optimality results presented
in \cite{wir}. In view of the equivalence between LU, LU+anc, LO and LOCC protocols for two--qubit systems, this solves the
problem of time--optimal, two--qubit Hamiltonian simulation under any of these classes of operations.

\vspace{5mm}

The structure of the paper is as follows. In section II we introduce some known results. Section III, IV and V present results ($i$), ($ii$) and ($iii$) respectively. Section VI contains some conclusions and appendices A and B discuss some technical aspects of sections III and V.

\section{Preliminaries}

We start by reviewing some background material from Ref. \cite{wir}, of which the present work can be regarded as an extension.

\subsection{Non--local Hamiltonian simulation and classes of operations}

Recall that the aim of non--local Hamiltonian simulation is, given a set of
systems that interact according to Hamiltonian $H$ for time $t$
and a class $C$ of local control operations, to be able to produce an
evolution $e^{-iH't'}$ for the systems, where $H'$ and $t'$ are
the simulated Hamiltonian and the simulated time. [We take
$\hbar \equiv 1$ along the paper]. 

As mentioned above, one can consider several classes of operations to assist in the
simulation, including LU, LU+anc, LO and LOCC. As in \cite{wir}, we make two basic assumptions: ($i$) these additional operations can be implemented very fast compared to the time scale of the Hamiltonian $H$ (we actually consider the setting in which they can be performed {\em instantaneously} and thus characterize the fast control limit); ($ii$) these operations are a cheap resource, so that optimality over simulation
protocols is defined only in terms of the ratio $t'/t$, that is,
in terms of how much time $t'$ of evolution according to $H'$
can be produced by using $H$ for a time $t$. Another interesting parameter characterizing simulations, that we do not analyze here, would be some measure of the complexity of the simulation, that is of the number of control operations that are performed.

We also note that the inclusions between classes of operations, LU $\subset$ LU+anc $\subset$ LO $\subset$ LOCC, imply relations between the sets of achievable simulations and time efficiencies. For instance, since LOCC simulation protocols contain all LU
simulation protocols, we expect LOCC
protocols to be more powerful than LU protocols.

\subsection{Infinitesimal--time simulations}

The maximal simulation factor $s(t')\equiv t'/t$ when simulating $e^{-iH't'}$ by using $H$ for time $t$ may depend on $t'$.
However, we are ultimately interested in characterizing the non-local properties of interaction Hamiltonians,
irrespective of interaction times.
A sensible way to proceed is by considering the worst case
situation, namely the time $t'$ for which the optimal ratio
$s(t')$ achieves its minimal value. This occurs for an
infinitesimal time $t'$. That is, simulations of $H'$ for a time
such that $||H't'||<< 1$ are, comparatively, the most expensive in
terms of the required time $t$ of interaction $H$. The reason is
that, ($i$) simulations for an infinitesimal time are a
particular case of simulation, providing an upper bound for the
minimum of $s(t')$, and ($ii$) any finite-time simulation ---or
{\em gate synthesis}--- can be achieved, maybe not optimally,
by concatenating infinitesimal--time simulations.

We shall denote $s_{H'|H}$ the limit $\lim_{t'\rightarrow 0}
s(t')$, and call it the simulation factor of $H'$ with $H$ [$s_{H'|H}$ corresponds to the inverse of the time overhead $\mu$ of Ref. \cite{Woc}, that is $s_{H'|H} = \mu^{-1}$.]
Then, apart from quantifying the time efficiency in 
infinitesimal simulations, $s_{H'|H}$ has also two other
meanings:
\begin{itemize}

\item $T'/s_{H'|H}$ upper bounds the time $T$ of use of $H$
needed to perform the unitary gate $e^{-iH'T'}$, for any $T'$ ({\em gate simulation} or {\em gate synthesis \cite{intcost}});

\item $s_{H'|H}$ is the optimal time efficiency in {\em dynamics simulation}. That is, $s_{H'|H}$ is the maximal achievable ratio $T'/T$, where $T$ is the time of $H$ required to
simulate the {\em entire} evolution of a system according to $e^{-it'H'}$, where $t'$ runs from $0$ to $T'$ \cite{simu}.

\end{itemize}

In an abuse of notation, we shall refer to condition $||H't'|| << 1$ as the small time limit, of which $\O(t') ~[$or $\O(t)]$ will denote first order corrections.

\subsection{Optimal and efficient simulations}

For any class $C \in \{$LU, LU+anc, LO, LOCC$\}$ of the above operations and in the small time
limit, the space of achievable evolutions using Hamiltonian $H$
and operations $C$ turns out to be convex. Then the following
two problems,

\vspace{2mm}

\noindent
{\bf P1:} {\it Given any $H$ and $H'$, determine when $H'$ can
be efficiently (i.e., $t'=t$) simulated with $H$ for
infinitesimal times, denoted}
\be
H' \geq_C H;
\ee

\vspace{2mm}

\noindent
{\bf P2:} {\it Given any $H$ and $H'$, determine the simulation
factor $s_{H'|H}$;}

\vspace{2mm}

\noindent
are equivalent, since $s_{H'|H}$ is nothing but the greatest $s$
such that $sH'$ can be efficiently simulated by $H$, that is, such that $sH' \geq_C H$.

\subsection{Equivalence of LO and LU+anc protocols}

The simulation of non--local Hamiltonians using LO and that using LU+anc are equivalent (see \cite{wir} for details), in that any protocol based on LO can be replaced with another one that uses only LU+anc and that has the same time efficiency. The ultimate reason for this equivalence is that even if LO provide, through measurement outcomes, information that can be used to decide on posterior local manipulations, this information cannot be transmitted to the other parties [unless the interaction itself is used for this purpose, but this leads to null efficiency $t'/t$ when $t\rightarrow 0$]; then, unitarity of the simulated evolution implies that each party is effectively applying a trace--preserving local operation on its subsystem, and this can always be achieved using only LU+anc. 

The previous situation changes when classical communication is allowed between the parties, because then they can coordinate their manipulations. In spite of this fact, CC does not help in Hamiltonian simulation, as we move to discuss next.

\section{Equivalence of LOCC and LU+anc protocols}

In this section we show that any protocol for non--local Hamiltonian simulation based on LOCC can be replaced with another one based only on LU+anc and having the same time efficiency. This result, valid for infinitesimal--time simulations on arbitrary $N$--particle systems, brings an important simplification to the general problem of non--local Hamiltonian simulation, since it implies the equivalence of LOCC, LO and LU+anc protocols.

We first describe in detail a most general protocol for Hamiltonian simulation using LOCC. Then we show ---through an argument that exploits the fact that entanglement only decreases under LOCC--- that any such protocol can be replaced with another one using only LU+anc. The key point of the proof is to assume that one of the systems is initially entangled with an auxiliary system $Z$, and to realize that a non--trivial measurement (i.e., a measurement not equivalent to some local unitary transformation) on the system would partially destroy this entanglement in an irreversible way. Since we are simulating a unitary process on the systems (which should preserve the entanglement between those and $Z$), all local measurements must be trivial, and can be replaced with unitary transformations.

\subsection{Hamiltonian simulation using LOCC}

For clearness sake we will perform most of the analysis in the
simplest non--trivial case, that involving only two qubits,
because this already contains all the ingredients of the general
$N$-particle setting. Let us consider, then, that qubits $A$ and $B$, with Hilbert spaces $\H_A$ and $\H_B$, interact according to $H$ for an overall time $t$, and that,
simultaneously, they are being manipulated locally. 

\subsubsection{Local manipulation}

The most general local operation on, say, qubit $A$ can be achieved \cite{book} by ($i$) appending to $A$ an ancillary system $A'$ in some blank state $\ket{0_{A'}} \in \H_{A'}$; ($ii$) performing a unitary transformation $U$ on $\H_{AA'} = \H_A\otimes \H_{A'}$; ($iii$) performing an orthogonal measurement on a factor space $\H_{meas}$ of the total Hilbert space $\H_{AA'} = \K\otimes \H_{meas}$, given by projection operators $\{P^{\alpha}\}$; and ($iv$) tracing out a factor space $\T^{\alpha}$ of $\H_{AA'}= \H_{out}^{\alpha}\otimes \T^{\alpha}$, where $\H^{\alpha}_{out}$ and $\T^{\alpha}$ may depend on the measurement outcome $\alpha$ of step ($iii$). Under ($i$)--($iv$) the initial state $\ket{\phi_A}$ of qubit $A$ transforms with probability $p_{\alpha}$ according to
\bea
&&\proj{\phi_A} \rightarrow \E_{\alpha}(\proj{\phi_A}) =\nonumber\\
&&\frac{1}{p_{\alpha}} \tr_{\T^{\alpha}} \left[P^{\alpha}U\left( \proj{\phi_A}\otimes\proj{0_{A'}} \right) U^{\dagger}P^{\alpha}\right],
\label{eq:ioio}
\eea
where $p_{\alpha} \equiv  \tr [P^{\alpha} U ( \proj{\phi_A}\otimes\proj{0_{A'}} ) U^{\dagger}P^{\alpha}]$. We can introduce operators $M^{\alpha}_i: \H_A \rightarrow \H_{out}^{\alpha}$,
\be
M_i^{\alpha} \equiv \bra{i^{\alpha}}P^{\alpha}U\ket{0_{A'}},
\ee
where $\{\ket{i^{\alpha}}\}$ is an orthonormal basis of $\T^{\alpha}$. Then Eq. (\ref{eq:ioio}) can be rewritten as
\be
\E_{\alpha}(\proj{\phi_A}) = \frac{1}{p_{\alpha}} \sum_i M^{\alpha}_i\proj{\phi_A}M^{\alpha\dagger}_i.
\ee
Now, since in our case the eventual result of this manipulation must be a unitary evolution, we are interested in transformations $\E_{\alpha}$ that map pure states into pure states, that is, such that can be implemented by just one operator $M^{\alpha}:\H_A \rightarrow \H_{out}^{\alpha}$,
\be
\proj{\phi_A} \rightarrow \frac{1}{p_{\alpha}}  M^{\alpha}\proj{\phi_A}M^{\alpha\dagger},
\ee
$p_{\alpha} = tr [M^{\alpha}\proj{\phi_A}M^{\alpha\dagger}]$. Therefore the effect of the local manipulation on qubit $A$ is a generalized measurement $\M$ that, with probability $p^{\alpha}$, maps the state of $A$ into a state supported on $\H_{out}^{\alpha}$,
\be
\ket{\phi_A} \rightarrow  \frac{1}{\sqrt{p^{\alpha}}} M^{\alpha}\ket{\phi_A},
\ee 
and produces classical information $\alpha$. The measurement operators $\{M^{\alpha}\}$ characterizing $\M$ satisfy $\sum_{\alpha} M^{\alpha \dagger}M^{\alpha} = I_{A}$.

More generally, in a simulation protocol measurement $\M$ may depend on some previous information $\beta$, in which case we write $\M^{\beta}$. In addition, the corresponding measurement operators $\{M^{\alpha,\beta}\}$ may map states from a two--dimensional subspace $\H_{in}^{\beta} \subset \H_{AA'}$ into another two dimensional subspace $\H_{out}^{\alpha, \beta}\subset \H_{AA'}$ that depends both on the measurement outcome $\alpha$ and on the previous information $\beta$, that is
\be
M^{\alpha,\beta}:H_{in}^{\beta}\rightarrow H_{out}^{\alpha, \beta}.
\ee
In the following, a series of measurements $\M$ will be concatenated, in such a way that the {\em out--subspace} $\H_{out}$ for a given measurement is related to the {\em in--subspace} $\H_{in}$ for the next one. 

We consider that a sufficiently large ancillary system $A'$ in a pure state has been initially appended to qubit $A$ so that it provides at once the extra degrees of freedom needed to perform all generalized measurements $\M$ on $A$. Finally, all the above considerations apply also to qubit $B$, to which an ancillary system $B'$ is appended.

\subsubsection{LOCC simulation protocol}

A LOCC protocol for simulating $e^{-it'H'}$ by $H$ for time $t$ is characterize by a partition $\{t_1, t_2,...,t_n\}$ of $t$, 
where $t_i\geq 0$, $\sum_i t_i = t$, and a series of local measurements,
$\{(\M_{0},\N_{0}),(\M_{1}^{\alpha_1},\N_{1}^{\alpha_1}), ...,
(\M_{n}^{\alpha_n},\N_{n}^{\alpha_n}) \}$. The protocol runs as follows: 

\begin{enumerate}

\item The simulation begins with measurements $\M_{0}$ and $\N_{0}$ being performed on $A$ and $B$, respectively. These map the original state of $AB$ into a state supported on some subspace of $AA'BB'$.
\item Then the two qubits $A$ and $B$ are left evolve according to $H$ for a time $t_1$.
\item After that, measurements $\M_{1}^{\alpha_1}$ and
$\N_{1}^{\alpha_1}$ are performed. Here, index $\alpha_1$
indicates that the measurements being performed after time $t_1$
may depend on the outcomes of measurements $\M_{0}$ and $\N_{0}$.
\item Again, the measurements are followed by an evolution, for time
$t_2$, of $A$ and $B$ according to $H$, and the protocol
continues in an iterative fashion. 
\item In step $k$, qubits
$A$ and $B$ are first left evolve according to $H$ for a time
$t_k$ and then measurements $\M_{k}^{\alpha_k}$ and
$\N_{k}^{\alpha_k}$ ($\alpha_k$ denoting again a possible dependence on the outcome of any previous measurement) are locally performed in $AA'$ and $BB'$.
\item The protocol finishes after measurements $\M_{n}^{\alpha_n}$ and
$\N_{n}^{\alpha_n}$ have been performed. These last measurements must leave the two--qubit system $AB$ in a pure state (that is, uncorrelated from systems $A'B'$, that are traced out). 
\end{enumerate}

Thus, the two--qubit system $AB$ is initially in some state $\ket{\psi}$, becomes entangled with the ancillas $A'$ and $B'$ during the manipulations describes above, but ends up in the state $e^{-iH't'}\ket{\psi}$ after time $t$.

Note that the protocol described above
has a tree structure, starting with a preestablished couple of
local manipulations and ending up at the extreme of a branch
characterized by the outcomes of all (conditioned) local
operations performed during the time interval $t$.  We move now to characterize one of these branches.

\begin{figure*}
\includegraphics{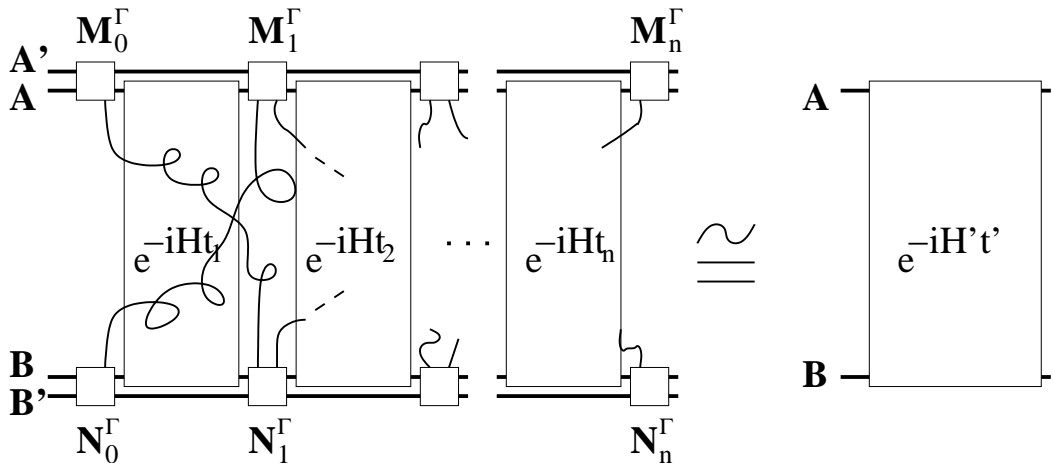}
\caption{\label{fig} Schematic representation of a Hamiltonian simulation protocol using LOCC. The unitary evolution of the composite system $AB$ according to $H$ and for a time $t=\sum_i t_i$ is interspersed with local measurements $\M_k^{\alpha_k}$ [on systems $AA'$] and $\N_k^{\alpha_k}$ [on systems $BB'$] to obtain a unitary evolution of $AB$ according to $H'$ and for a time $t'$. Here $\alpha_k$ indicates the local measurements performed at step $k$, which  may depend on the outcome of all previous steps thanks to the ability to communicate classical between the systems (winding lines). In the figure we have replaced the superscripts $\alpha_k$ with $\Gamma$, $\Gamma$ denoting a particular branch of the protocol (cf. Eq. (\ref{eq:locc})). Thus, in branch $\Gamma$ measurement operators $M_k^\Gamma$ [corresponding to measurement $\M_k^{\Gamma}$] and $N_k^\Gamma$ [corresponding to measurement $\N_k^{\Gamma}$] transform systems $AA'$ and $BB'$ at step $k$ of the protocol.}
\end{figure*}

\subsubsection{One branch of the protocol}

Let us suppose we run the simulation once. This corresponds to some given branch of the protocol, that we label $\Gamma$, and have represented in the figure. Branch $\Gamma$ is characterized by a series of
measurement operators $\{(M_{0}^{\Gamma}, N_{0}^{\Gamma}), ...,
(M_{n}^{\Gamma}, N_{n}^{\Gamma})\}$, where the superindices $\alpha_k$ containing the information that characterizes the branch have been replaced with $\Gamma$ to simplify the notation. Recall that the aim of the protocol is to achieve an evolution according to $e^{-iH't'}$. Therefore, for any initial vector
$\ket{\psi}$ of the two--qubit system $AB$, the measurement operators $\{(M_{k}^{\Gamma}, N_{k}^{\Gamma})\}_{k=0}^n$ must obey
\bea
&& \sqrt{p_{\Gamma}} e^{-iH't'} |\psi\ra = ( M_n^{\Gamma}\otimes N_n^{\Gamma})
e^{-it_nH} \cdots \nonumber \\
&&~~~~~~~~~~~\cdots( M_1^{\Gamma}\otimes N_1^{\Gamma})
e^{-it_1H} 
(M_0^{\Gamma}\otimes N_0^{\Gamma} ) |\psi\ra,
 \label{eq:locc}
\eea
where $p_{\Gamma}$ denotes the probability that branch $\Gamma$ occurs in the protocol. Eq. (\ref{eq:locc}) is the starting point for the rest of the analysis in this section.

\subsection{LOCC protocols are as efficient as LU+anc protocols for infinitesimal time simulations}

As discussed in the introduction, we are interested here in simulations for an infinitesimal simulation time $t$. In this
regime Eq. (\ref{eq:locc}) significantly simplifies, because we
can expand the exponentials to first order in $t$ (or equivalently, in $\{t_k\}$ and $t'$), thereby
obtaining an equation which is linear both in $H$ and $H'$. In addition, if
$t$ is small then qubits $A$ and $B$ interact only ``a little
bit''. In what follows we will use this fact to prove the main
result of this section, namely that all the measurement
operators $\{M_k^{\Gamma}, N_k^{\Gamma}\}_{k=0}^n$ in Eq. (\ref{eq:locc}) must be, up
to negligible corrections, proportional to unitary operators in
some corresponding relevant supports. This will eventually imply 
that LU+anc protocols can already simulate any
evolution $e^{-iH't'}$ achievable in a LOCC protocol. 

We note that this result is not valid for the interconversion of non--local gates \cite{gates}. There the systems are allowed to interact according to a finite gate (e.g. a C--NOT), and thus accumulate some finite amount of entanglement (e.g. an {\em ebit}) in the ancillary systems, that can be used, together with LOCC, to perform some new non--local gate (e.g. through some teleportation scheme).

\subsubsection{LOCC protocols for infinitesimal--time simulations}

We define a series of operators $M_k$ and $M_k'$ by
\bea
&&M_k \equiv M_n^{\Gamma}\cdots M_{k}^{\Gamma},~~~~~~k=1,\cdots,n,\nonumber\\
&&M_k' \equiv M_{k-1}^{\Gamma}\cdots M_0^{\Gamma},~~~~~~k=1,\cdots,n, \nonumber\\
&&M_0 \equiv M_n^{\Gamma}\cdots M_{0}^{\Gamma}, 
\eea
and also an analogous series of operators $N_k,N_k'$ and $N_0$. Notice that operator $M_k'$ describes a concatenation of all local measurements in branch $\Gamma$ performed from the beginning of the protocol and up to step $k-1$ on the state initially supported on $\H_{AA'}$, while $M_k$ collects the manipulations that will be performed from step $k$ until the end of the protocol. In the small time regime, we can expand the exponentials in Eq. (\ref{eq:locc}) as a series in $t_k$ and $t'$ to obtain, up to second order corrections $\O(t^2)$,
\bea
\sqrt{p_{\Gamma}} \left( I_{AB} - istH' \right) \ket{\psi}= ~~~~~~~~~~~~~~~~\label{eq:inflocc0} \\ 
\left( M_0\otimes N_0 - i t\sum_{k=1}^n p_k
 (M_k\otimes N_k) H (M_k'\otimes N_k')\right)\ket{\psi},\nonumber
\eea
where we have introduced probabilities $p_k\equiv t_k/t$ and the efficiency factor $s\equiv t'/t$ of the branch, so that all times are expressed in terms of $t$.

This equation indicates that
\be
M_0\otimes N_0 \ket{\psi}=  \sqrt{p_{\Gamma}} I_{AB} \ket{\psi}+ \O(t),
\label{eq:identity}
\ee
for any two--qubit state $\ket{\psi}$, from which it follows that the probability $p_{\Gamma}$ that branch $\Gamma$ occurs cannot depend on $\ket{\psi}$ up to $\O(t)$ corrections, also that both $M_0$ and $N_0$ must be proportional to the identity operator in $H_A$ and $H_B$, 
 \bea
 M_0 &=& \sqrt{p_{\Gamma}}q I_{A} + \O(t),\\
 N_0 &=& ~~q^{-1} I_{B} + \O(t),
 \eea
where $q$ is some positive parameter.
Notice that the order $t$ corrections in Eq. (\ref{eq:identity}) correspond to local terms, that is, to operators of the form $t(I_A\otimes O_B + O'_A\otimes I_B)$, and thus are irrelevant to this discussion \cite{neglect}. In what follows we neglect these local terms for clearness sake. Bearing this remark and Eq. (\ref{eq:identity}) in mind, we rewrite Eq. (\ref{eq:inflocc0}) as the operator equation
\bea
&& \sqrt{p_{\Gamma}} \left(I_{AB} - istH' \right)= \label{eq:inflocc}
\\ &&
\sqrt{p_{\Gamma}} I_{AB}
- i t\sum_{k=1}^n p_k
 (M_k\otimes N_k) H (M_k'\otimes N_k') + \O(t^2).\nonumber
\eea
That is,
\be
\sqrt{p_{\Gamma}}sH' = \sum_{k=1}^n p_k (M_k\otimes N_k) H (M_k'\otimes N_k') + \O(t),
\label{eq:semihami}
\ee
where, because of Eq. (\ref{eq:identity}), some other constrains apply. More precisely, if $M_k'$ and $N_k'$ are given by
\bea
M_k' = \sqrt{p_{\Gamma}}q(\ket{\mu_0^k}\bra{0_A} + \ket{\mu_1^k}\bra{1_A})\nonumber\\
N_k' = ~~q^{-1}(\ket{\nu_0^k}\bra{0_B} + \ket{\nu_1^k}\bra{1_B}), \label{eq:Mk}
\eea
where $\{\ket{i_A}\}$ and $\{\ket{i_B}\}$ are orthonormal basis of $\H_A$ and $\H_B$ and $\{\ket{\mu_i^k}\in \H_{AA'}\}$ and $\{\ket{\nu_i^k}\in \H_{BB'}\}$ are arbitrary vectors, not necessarily normalized, then  $M_k$ and $N_k$ must fulfill 
\bea
M_k &=& \ket{0_A}\bra{\tilde{\mu}_0^k} + \ket{1_A}\bra{\tilde{\mu}_1^k} + \O(t),\nonumber\\
N_k &=& \ket{0_B}\bra{\tilde{\nu}_0^k} + \ket{1_A}\bra{\tilde{\nu}_1^k}+\O(t), \label{eq:Mk'}
\eea
where $\{\ket{\tilde{\mu}_i^k}\}$ is the biorthonormal basis of $\{\ket{\mu_i^k}\}$ (in the subspace spanned by $\{\ket{\mu_i^k}\}$), that is $\braket{\mu_i^k}{\tilde{\mu}_j^k} = \delta_{ij}$, and  similarly $\{\ket{\tilde{\nu}_i^k}\}$ is the biorthonormal basis of $\{\ket{\nu_i^k}\}$, so that $M_0\otimes N_0 = (M_kM_k')\otimes(N_kN_k')$ fulfills Eq. (\ref{eq:identity}).

Now, going back to the measurement operators $M_k^{\Gamma}$, we can expand them as
\bea
M_0^{\Gamma} &=& \sqrt{p_{\Gamma}}q\left(~ \ket{\mu_0^1}\bra{0_A}+\ket{\mu_1^1}\bra{1_A} ~\right)\nonumber\\
M_k^{\Gamma} &=& \ket{\mu_0^{k+1}}\bra{\tilde{\mu}_0^{k}}+ \ket{\mu_1^{k+1}}\bra{\tilde{\mu}_1^{k}} ~~~~~~k=1,\cdots, n\!-\!1\nonumber\\
M_n^{\Gamma} &=& \ket{0_A}\bra{\tilde{\mu}_0^{n}} + \ket{1_A}\bra{\tilde{\mu}_1^{n}}+ \O(t) \label{eq:operators}
\eea
and similarly for the $N_k^\Gamma$.

\subsubsection{Unitarity and conservation of entanglement}

We carry on this analysis by focusing our attention only on the operations performed on
systems $AA'$. We will show that operators
$M_k$ and $M_k'$ can be replaced with operators proportional to $\bra{0_{A'}}U_k$ and $U_k^{\dagger}\ket{0_{A'}}$, where $U_k$ is a unitary matrix acting on $H_{AA'}$. We will use the fact that the
protocol must be able to keep the entanglement of $A$ with
another system $Z$.

Let us suppose, then, that qubit $A$ is entangled with a distant
qubit $Z$, with the maximally entangled vector 
\be
\frac{1}{\sqrt{2}}(\ket{0_{A}}\otimes \ket{0_{Z}} + \ket{1_{A}}\otimes \ket{1_{Z}}) \label{eq:entangled}
\ee
describing the pure state of $AZ$. Any
unitary evolution of qubits $A$ and $B$ preserves the amount of
entanglement between qubit $Z$ and qubits $AB$. In particular, if the unitary evolutions according to $H$ are infinitesimal, then up to $\O(t)$ corrections qubit $Z$ must be still in a maximally entangled state with $A$ after the
simulated evolution $e^{-istH'}$. This sets very strong
restrictions on the kind of measurements that can be performed on $A$ during
the simulation protocol. If during the $k^{th}$ measurement in branch $\Gamma$ part of the entanglement is destroyed, then the simulation protocol necessarily fails with some probability, because the destroyed entanglement can not be deterministically recovered. Indeed, even if subsequent measurement operators in branch $\Gamma$ would be able to restore the entanglement and so obey Eq. (\ref{eq:locc}), another branch $\Gamma'$ diverging from $\Gamma$ after the $k^{th}$ must necessarily fail to recover the entanglement (recall the monotonically decreasing character of entanglement under LOCC, see e.g. \cite{VJN}) and thus with some probability the protocol must fail to simulate the unitary evolution \cite{robustness}.

Let us see the effect of this restrictions on the first measurement operator $M_0^{\Gamma}$ in Eq. (\ref{eq:operators}). It transforms the initial entangled state into a new state proportional to
\be
\ket{\mu_0^1}\otimes \ket{0_{Z}} + \ket{\mu_1^1}\otimes \ket{1_{Z}},
\ee
which remains maximally entangled if and only if $||\ket{\mu_0^1}|| =||\ket{\mu_1^1}|| \equiv r_1$ and $\braket{\mu_0^1}{\mu_1^1}=0$. But this are precisely the conditions for $M_1' (=M^{\Gamma}_0)$ to be proportional to a unitary operator from $\H_A$ to the out--space $\H_{out}^0$ spanned by $\{\ket{\mu_i^1}\}$ or, equivalently, to an isometry from $\H_A$ to $\H_{AA'}$. Thus, we can write
\be
M_1' = r_1 U^{\dagger}_1\ket{0_{A'}},
\ee
where
\bea
U_1^{\dagger} \equiv \frac{\ket{\mu_0^1}}{r_1}\bra{0_A0_{A'}} + \frac{\ket{\mu_1^1}}{r_1}\bra{1_A0_{A'}}\nonumber\\
            +\sum_{l=1}^{d_{A'}-1} \ket{\xi_{l,0}}\bra{0_Al_{A'}} + \ket{\xi_{l,1}}\bra{1_Al_{A'}} \nonumber
\eea
is some unitary operation defined on $\H_{AA'}$. Here $d_{A'}$ is the dimension of $\H_{A'}$ and $\{\ket{\xi_{l,0}},\ket{\xi_{l,1}} \}_{l=1}^{d_{A'}}$ is a set of irrelevant vectors that together with $\ket{\mu_0^1}/r_1$ and $\ket{\mu_1^1}/r_1$ form an orthonormal basis of $\H_{AA'}$. Eq. (\ref{eq:Mk'}) implies that, in addition,
\be
M_1 = \frac{\sqrt{p_{\Gamma}}q}{r_1}\bra{0_{A'}} U_1.
\ee

This characterization in terms of a unitary transformation can now be easily extended to the rest of operators $M_k$ and $M_k'$. We use induction over $k$. We already have that the characterization works for $k=1$. Suppose it works for some $k-1$, that is, in the decomposition Eq. (\ref{eq:Mk}) for $M'_{k-1}$ we have $||\ket{\mu^{k-1}_0}||=||\ket{\mu^{k-1}_1}||$ and $\braket{\mu^{k-1}_0}{\mu^{k-1}_1}=0$. This means that after the $(k-1)^{th}$ measurement in branch $\Gamma$, the initial state of Eq. (\ref{eq:entangled}) becomes a state proportional to
\be
\ket{\mu_0^{k-1}}\otimes \ket{0_{Z}} + \ket{\mu_1^{k-1}}\otimes \ket{1_{Z}} + \O(t),
\ee
where the $\O(t)$ corrections are due to evolutions of $AB$ according to $H$ for a time of order $t$, which slightly entangle $B$ with $AZ$. Then, preservation of entanglement during the $k^{th}$ measurement (implemented by operator $M^{\Gamma}_{k-1}$) requires that also $||\ket{\mu_{k}^0}||=||\ket{\mu_{k}^1}||\equiv r_{k}$ and $\braket{\mu_{k}^0}{\mu_{k}^1}=0$, and therefore
\bea
M_{k}'&=& r_{k} U_{k}^{\dagger} \ket{0_{A'}} + \O(t), \nonumber \\
M_{k} &=& \frac{\sqrt{p_{\Gamma}}q}{r_k}\bra{0_{A'}} U_k +\O(t),
\eea
for some unitary transformation $U_k$ acting on $\H_{AA'}$.

The same argument leads to expressing the operators $N_k$ and $N_k'$ in terms of unitary transformations $V_k$ acting on $\H_{BB'}$ as
\bea
N_{k}'&=& s_{k} V_{k}^{\dagger} \ket{0_{B'}} + \O(t), \nonumber \\
N_{k} &=& \frac{1}{r_k q}\bra{0_{B'}} V_k +\O(t).
\eea
Therefore Eq. (\ref{eq:semihami}) finally reads, up to $\O(t)$ corrections that vanish in the $t\rightarrow 0$ or fast control limit, 
\be
sH' = \sum_k \!p_k \!\bra{0_{A'}0_{B'}}(U_k\otimes V_k) H (U^{\dagger}_k\otimes V^{\dagger}_k) \ket{0_{A'}0_{B'}}.
 \label{eq:convexsum}
\ee

\subsubsection{Equivalence between LOCC and LU+anc protocols}

The set $S_{H}^{LOCC}$ of non--local Hamiltonians that can be efficiently simulated
by $H$ and LOCC is {\em convex}: if $H$ can efficiently simulate $H_1$
and $H_2$, then it can also efficiently simulate the Hamiltonian $pH_1+(1-p)H_2$.
Indeed, we just need to divide the infinitesimal time $t$ into
two parts and simulate $H_1$ for time $pt$ and then $H_2$ for
time $(1-p)t$. The resulting Hamiltonian is precisely the above
average of $H_1$ and $H_2$. Thus, in order to characterize the
convex set $S_{H}^{LOCC}$, we can focus on its {\em extreme points}. Notice that the previous convexity argument also holds for the set $S_{H}^{LU+anc}$ of Hamiltonians that can be efficiently simulated with LU+anc, so that $S_{H}^{LU+anc}$ is also convex. Recall also that $S_{H}^{LU+anc} \subset S_{H}^{LOCC}$.

Now, Eq. (\ref{eq:convexsum}) says that all points in $S_{H}^{LOCC}$ can be
obtained as a convex combination of terms of the form
\be
 \bra{0_{A'} 0_{B'}} (U\otimes V )H (U^{\dagger}
 \otimes V^{\dagger})  \ket{0_{A'} 0_{B'}}.
\label{eq:extreme}
\ee
In addition, in appendix A we show that any such a term can be
obtained in a simulation protocol using LU+anc. It follows that ($i$) any extreme point of $S_{H}^{LOCC}$
is of the form (\ref{eq:extreme}), and that ($ii$) any extreme point of $S_H^{LOCC}$ belongs to $S_{H}^{LU+anc}$, so that $S_{H}^{LU+anc} = S_{H}^{LOCC}$. 
This finishes
the proof of the fact that infinitesimal time simulations using
LOCC can always be accomplished using LU+anc. 

Summarizing, we have seen that any (rescaled) two-qubit Hamiltonian
$sH'$ achievable in branch $\Gamma$ of our LOCC-simulation
protocol (cf. Eq. (\ref{eq:convexsum})) can also be achieved, with the same time efficiency, by just using local unitary transformations and ancillas as extra resources.
It is now straightforward to generalize the above argument to
$N$ systems, each one having two or more levels, thereby
extending the equivalence of LOCC and LU+anc protocols to
general multiparticle interactions. Indeed, for any $d$-level
system involved in the simulation, we just need to require that
its entanglement with some remote, auxiliary $d$-level system be
preserved, and we readily obtain that all measurements performed
during the simulation protocol can be replaced with local
unitary operations. We thus can conclude, using the notation introduced in section II.B, that
\be
H' \geq_{LOCC} H \Leftrightarrow H' \geq_{LU+anc} H.
\ee

\section{LU+anc protocols are not equivalent to LU protocols}

The equivalence between infinitesimal--time simulations using
LOCC and LU+anc may be conceived as a satisfactory result.
On the one hand, it discards local measurements and classical communication as useful resources for the simulation of non--local Hamiltonians. This essentially says that in order to simulate Hamiltonian dynamics, we can restrict the external manipulation to unitary operations, possibly involving some ancillary system. In this way the set of interesting simulation protocols has been significantly simplified. On the other hand, it
is reassuring to see that, despite the diversity of classes of
operations that we may use as a criterion to characterize the
non-local properties of multiparticle interactions, most of
these criteria (LOCC, LO and LU+anc) yield an equivalent 
classification and quantification. In other words, we do not
have to deal with a large number of alternative
characterizations. We shall show here, however, that simulation
using only LU, that is, without ancillas, is not equivalent to
that using LU+anc.

The reason for this inequivalence is the following. Consider a multi--partite Hamiltonian of the form $H_A\otimes H_{BC\cdots}$, where $H_A$ acts on a $d$ dimensional space $\H_A$ and $H_{BC\cdots}$ acts on $\H_B\otimes \H_C \cdots$. In the presence of an ancilla $\H_{A'}$, LU can be used so that operator $H_A$ acts on some $d$ dimensional {\em factor} space $\K$ of $\H_{AA'}$ ($\H_{AA'} = \K\otimes \K'$). The net result is an effective Hamiltonian acting on $\H_A$. As the following examples show, some of these effective Hamiltonians can not be achieved (at least with the same time efficiencies) by using only LU. 

\subsection{LU+anc protocols versus LU protocols}

In the previous section we saw that, in the fast control limit, the extreme points of the convex set  $S_H^{LU+anc}$ of bipartite Hamiltonians that can be
efficiently simulated with $H$ using LU+anc, [equivalently, those of the set $S_H^{LOCC}$] are, up to local terms, of the form
\be
\E(H) \equiv \bra{0_{A'} 0_{B'}} U\!\otimes\! V (H\otimes I_{A'B'})
U^{\dagger}
 \!\otimes\! V^{\dagger}  \ket{0_{A'} 0_{B'}}
\label{eq:extreme2}
\ee
(an analogous expression holds for the multi--partite case). Notice that in Eq. (\ref{eq:extreme2}) we have replaced operator $H$ of Eq. (\ref{eq:extreme}) with $H\otimes I_{A'B'}$ to make more explicit that ancillas are being used.
 
Can all simulations of this type be achieved by using only LU? The most general simulation that can be achieved from $H$ and by LU reads (see \cite{wir} for more details)
\bea
&&\sum_k p_k u_k\otimes v_k Hu_k^{\dagger}\otimes v_k^{\dagger} \nonumber\\
&+& m\otimes I_B + I_A \otimes n + aI_{AB},
\label{eq:uniconj}
\eea
where $\{p_k\}$, $\sum_k p_k=1$, is a probability distribution,  $\{u_k\}$ and $\{v_k\}$ are local unitaries acting on $A$ and $B$, $m$ and $n$ are
self-adjoint, trace-less operators and $a$ is a real constant. The previous question translates then into whether for any $U$ and $V$ in Eq. (\ref{eq:extreme2}), we can find a set $\{p_k, u_k, v_k\}$, $m$, $n$ and $a$ such that Eq. (\ref{eq:uniconj}) equals $\E(H)$ in (\ref{eq:extreme2}).

In \cite{wir} it was shown that, in the particular case of
two-qubit systems, the previous conditions can always be
fulfilled. Next we shall show that this is sometimes not the case
for Hamiltonians of two $d$-level systems for $d>2$, and also
for Hamiltonians of more than two systems.

\subsection{Inequivalence between LU+anc and LU protocols}

\subsubsection{Example 1: two $d$-level systems ($d>2$)}

We first consider two $d$--level systems $A$ and $B$, $d>2$, that interact according to
\be
K\equiv P_{0}\otimes P_{0} + \sum_{i=1}^{d-1} P_{i}\otimes P_{i},
\ee
where $P_i\otimes P_j\equiv \proj{i_{A}}\otimes\proj{j_{B}}$. We will show that by means of LU+anc, Hamiltonian $K$ can be used to efficiently (that is, with unit efficiency factor $s$) simulate
\be
K' \equiv  P_0\otimes P_1 + \sum_{i=1}^{d-1} P_{i}\otimes P_{i}.
\ee
We will also show that $K'$ can not be efficiently simulated using only LU.

Let $A'$ be a $d$-level ancilla. We need a unitary transformation $U$ satisfying
\be
\bra{0_{A'}}U = \ket{0_{A}}\bra{1_{A}0_{A'}} + \sum_{i=1}^{d-1} \ket{i_{A}}\bra{i_{A}i_{A'}}.
\ee
As we discuss in appendix A, the transformation of a Hamiltonian $H$ acting on $AB$,
\be
\E(H)\equiv\bra{0_{A'}0_{B'}} U (H\otimes I_{A'B'})U^{\dagger}\ket{0_{A'}0_{B'}},
\ee
can be achieved using LU+anc [notice that this corresponds to choosing $V_{BB'}=I_{BB'}$ in Eq. (\ref{eq:extH'})]. In particular, this transformation takes any term of the form $P_i\otimes P_j$ into
\be
\E(P_i\otimes P_j)= \left\{ \begin{array}{cc} 0 & i = 0\\ (P_0+P_1)\otimes P_j & i=1\\ P_i\otimes P_j & i>1.
\end{array} \right.,
\ee
which in particular implies
\be
\E(K) = K'.
\ee
Now, if this simulation is to be possible with the same time efficiency by using only LU, then we must have, because of Eq. (\ref{eq:uniconj}),
\be K' = Q + m\otimes I_B + I_A \otimes n + aI_{AB},
\label{eq:igual}
\ee
where $Q \equiv \sum_{i=0}^{d-1} \sum_k p_k u_kP_iu_k^{\dagger}\otimes v_k P_i v_k^{\dagger} \geq 0$,
but this is not possible. Indeed, we first notice that, by taking the trace of this expression we obtain $a=0$, whereas by tracing out only system $B$ we obtain
\be
I = I + dm,
\ee
and thus $m=0$. Tracing out only system $A$ leads to
\be
2 P_1 + \sum_{i=2}^{d-1} P_i = I + dn,
\ee
so that $n = (-P_0+P_1)/d$ and condition (\ref{eq:igual}) becomes
\be
K'=P_0\otimes P_1 + \sum_{i=1}^{d-1} P_{i}\otimes P_{i} = Q + \frac{I_A}{d}\otimes (-P_0+P_1).
\ee
Then, recalling positivity of $Q$, we obtain the following contradiction
\bea
0 &=& \mbox{tr } [P_2\otimes P_1 K'] 
= \mbox{ tr } [P_2\otimes P_1 Q]~ \nonumber\\ 
&+& \mbox{ tr } [(P_2\otimes P_1) \left(\frac{I_A}{d}\otimes(-P_0+P_1)\right)]~\nonumber\\
&=& \mbox{ tr } [(P_2\otimes P_1) Q]~ + 1/d \geq 1/d.
\eea

Thus, for any $d>2$, we have explicitly constructed an example of LU+anc simulation for Hamiltonians acting on two $d$-level systems that can not be achieved using only LU. We recall, however, that for two--particle Hamiltonians, LU+anc and
LU protocols only differ quantitatively, for LU protocols are
able to simulate any bipartite Hamiltonian $H'$ starting from
any other $H$ with non-vanishing $s_{H'|H}$ \cite{Dod,wir,Woc2,Nie,Woc3}.

\subsubsection{Example 2: a $2\times 2\times 2$ composite system}

Let us consider now the simulation, for an infinitesimal time
$t$, of the three--qubit Hamiltonian 
\be
K' \equiv I\otimes \sigma_3 \otimes \sigma_3
\ee
by the Hamiltonian
\be
K \equiv \sigma_3 \otimes \sigma_3 \otimes \sigma_3,
\ee
where
\be
\sigma_3 \equiv \mattwo{1}{0}{0}{-1}.
\ee
This is possible, when allowing for LU+anc operations, by
considering the transformation $U$ acting on qubit $A$ and on a
one--qubit ancilla $A'$ in state $\ket{0_{A'}}$, where
\be
\bra{0_{A'}}U = \ket{0_A}\bra{0_A}\otimes\bra{0_{A'}}+ \ket{1_A}\bra{0_A}\otimes\bra{1_{A'}},
\ee
Indeed, we have that $\bra{0_{A'}}U (\sigma_3\otimes I_{A'})
U^{\dagger}\ket{0_{A'}} = I_{A}$, so that
\be
\bra{0_{A'}}UKU^{\dagger}\ket{0_{A'}}=K'.
\ee 
On the other hand it is impossible to simulate $K'$ by $K$ and LU, for it would
imply to transform $\sigma_3$ into $I$ through unitary mixing,
which is a trace-preserving operation. It is straightforward to
construct similar examples in higher dimensional systems, and
also with more than three systems.

We note that, as far as interactions involving more than two
systems are concerned, the inequivalence between LU+anc and LU
simulation protocols is not only quantitative, leading to
different simulation factors, but also qualitative. The last example
above shows that LU protocols can not be used to simulate
Hamiltonians that can be simulated using LU+anc and the same
interaction $H$.

\section{Optimal simulation of two-qubit Hamiltonians using LOCC}

In this last section we address the problem of optimal
Hamiltonian simulation using LU for the case of two-qubit
interactions. We recover the results of \cite{wir}, but through
an alternative, simpler proof, based on known results of
majorization theory ---and thus avoiding the geometrical
constructions of the original derivation \cite{wir}. The equivalence of
LOCC and LU+anc strategies presented in section III, together with
that of LU+anc and LU strategies for two-qubit Hamiltonians
proved in \cite{wir}, imply that these results are also
optimal in the context of LOCC, LO and LU+anc Hamiltonian simulation.

We start by recalling some basic facts. Any two--qubit
Hamiltonian $H$ is equivalent, as far as LU simulation protocols
are concerned, to its canonical form \cite{Dur,wir}
\be
H = \sum_{i=1}^3 h_i \sigma_i\otimes\sigma_i,
\label{eq:cano1}
\ee
where $h_1 \geq h_2 \geq |h_3| \geq 0$ and the operators
$\sigma_i$ are the Pauli matrices,
\bea
\sigma_1 \equiv \mattwo{0}{1}{1}{0},~\sigma_2 \equiv \mattwo{0}{-i}{i}{0},\sigma_3 \equiv \mattwo{1}{0}{0}{-1}.
\eea
A brief justification for this canonical form is as follows. Any two-qubit Hamiltonian 
\be
H_A\otimes I_B + I_A\otimes H_B + \sum_{ij} h_{ij} \sigma_i\otimes\sigma_j
\ee
can efficiently simulate (or be efficiently simulated by) its canonical form (\ref{eq:cano1}): on the one hand we can always use traceless operators $m$ and $n$ as in Eq. (\ref{eq:uniconj}) to remove (or introduce) the local operators $H_A$ and $H_B$; then the remaining operator $\sum_{ij} h_{ij} \sigma_i\otimes\sigma_j$ can be taken into the canonical form by means of one-qubit unitaries $u$ and $v$ such that $(u\otimes v)\sum_{ij} h_{ij} \sigma_i\otimes\sigma_j (u^{\dagger}\otimes v^{\dagger})$ is diagonal when expressed in terms of Pauli matrices. The coefficients $h_i$ in Eq. (\ref{eq:cano1}) turn out to be related to the singular values of the matrix $h_{ij}$.

Therefore we only need to study the conditions for efficient simulation between Hamiltonians which are in a canonical form. Let $\{\ket{\Phi_i}\}$ stand for the basis of maximally
entangled vectors of two qubits
\bea
\ket{\Phi_{1}} \equiv \frac{1}{\sqrt{2}} (\ket{01} + \ket{10}),~~ \ket{\Phi_{2}} \equiv \frac{1}{\sqrt{2}} (\ket{00} + \ket{11}),\nonumber\\
\ket{\Phi_{3}} \equiv \frac{1}{\sqrt{2}} (\ket{00} - \ket{11}),~~ \ket{\Phi_{4}} \equiv \frac{1}{\sqrt{2}} (\ket{01} - \ket{10}).
\label{eq:max}
\eea
Then $H$ can be alternatively expressed as
\be
H = \sum_{i=1}^{4} \lambda_i \proj{\Phi_i},
\label{eq:cano2}
\ee
where $\lambda_i$ are decreasingly ordered, real coefficients
fulfilling the constraint $\sum_{i} \lambda_i = 0$ (coming from
the fact that $H$ has no trace) and
\bea
\lambda_1 &=& h_1+h_2-h_3\\
\lambda_2 &=& h_1-h_2+h_3\\
\lambda_3 &=& -h_1+h_2+h_3\\
\lambda_4 &=& -h_1-h_2-h_3.
\eea

The most general simulation protocol using $H$ and LU leads to
\be
H' = \sum_k p_k u_k\otimes v_k H u_k^{\dagger}\otimes
v_k^{\dagger},
\label{eq:unimixing}
\ee
where we have assumed, without loss of generality, that $H'$ is also
in its canonical form, as in Eqs. (\ref{eq:cano1}) and
(\ref{eq:cano2}), with corresponding coefficients $h_i'$ and
$\lambda_i'$.

\subsection{Necessary and sufficient conditions for efficient simulation and optimal simulation factor}

Let us derive the necessary and sufficient conditions for $H$ to
be able to simulate $H'$ using LU and for infinitesimal
simulation times. Uhlmann's theorem \cite{Uhl} states that the
eigenvalues $\lambda'_i$ of operator $H'$ in Eq.
(\ref{eq:unimixing}), a unitary mixing of operator $H$, are
majorized by the eigenvalues $\lambda_i$ of $H$, that is
\bea
\lambda'_1 &\leq& \lambda_1,\nonumber\\
\lambda'_1 + \lambda'_2 &\leq& \lambda_1 + \lambda_2,\nonumber\\
\lambda'_1 + \lambda'_2 + \lambda'_3 &\leq& \lambda_1 + \lambda_2 + \lambda_3, \nonumber\\
\lambda'_1 + \lambda'_2 + \lambda'_3 + \lambda'_4 &=& \lambda_1 + \lambda_2 + \lambda_3 + \lambda_4, \label{eq:maj}
\eea
where the last equation is trivially fulfilled due to the fact
that $H$ and $H'$ are trace-less operators. Succinctly, we shall
write $\vec{\lambda'} \prec \vec{\lambda}$, as usual \cite{maj}.
In terms of the coefficients $h_i$ and $h'_i$ the previous
conditions read
\bea
h_1' &\leq& h_1,\nonumber\\ h_1' + h_2' - h_3' &\leq& h_1 + h_2
- h_3,\nonumber\\ h_1' + h_2' + h_3' &\leq& h_1 + h_2 + h_3,
\label{eq:omaj}
\eea
and correspond to the s(pecial)-majorization relation, $\vec{h'}
\prec_s \vec{h}$, introduced in Ref. \cite{wir}. Thus, we have
already recovered the necessary conditions of \cite{wir} for $H$ to
be able to {\em efficiently} simulate $H'$ in LU protocols
\cite{Bernstein} (and thus, since we are in the two--qubit case,
also in LOCC protocols).

In order to see that conditions (\ref{eq:maj}) [and thus
conditions (\ref{eq:omaj}) ] are also sufficient for efficient
LU simulation, we concatenate two other results of majorization
theory. The first one (see theorem II.1.10 of \cite{maj}) states
that $\vec{\lambda'}\prec \vec{\lambda}$ if and only if a doubly
stochastic matrix $m$ exists such that $\lambda_i'= \sum_j
m_{ij} \lambda_j$. The second result is known as Birkhoff's
theorem \cite{maj}, and states that the matrix $m$ can always be
written as a convex sum of permutation operators $\{P_k\}$, so
that
\be
\left( \begin{array}{c} \lambda_1' \\ \lambda_2' \\ \lambda_3' \\\lambda_4' \end{array} \right) = \sum_k p_k P_k \left( \begin{array}{c} \lambda_1 \\ \lambda_2 \\ \lambda_3 \\ \lambda_4 \end{array} \right).
\ee
This means that whenever conditions (\ref{eq:maj}) are fulfilled
we can obtain $H'$ from $H$ by using a mixing of unitary
operations $T_i$, where each $T_i$ permutes the vectors
$\{\ket{\Phi_i}\}$,
\be
H' = \sum_i p_i T_i H T_i^{\dagger}.
\ee
Then, all we still need to see is that all $4!=24$ possible
permutations of the vectors $\{\ket{\Phi_i}\}$ can be performed
through {\em local} unitaries $T_i$. Recall, however, that any
permutation $\sigma$, taking elements $(1,2,3,4)$ into
$(\sigma(1),\sigma(2),\sigma(3), \sigma(4))$, can be obtained by
composing (several times) the following three transpositions,
\bea
(1,2,3,4) \rightarrow (2,1,3,4),\\ (1,2,3,4) \rightarrow
(1,3,2,4),\\ (1,2,3,4) \rightarrow (1,2,4,3),
\eea
where each permutation affects two neighboring elements. The
corresponding three basic permutations of
$(\Phi_1,\Phi_2,\Phi_3,\Phi_4)$ can be easily obtained using LU.
Indeed, in order to permute $(\Phi_1,\Phi_2,\Phi_3,\Phi_4)$ into
\bea
(\Phi_2,\Phi_1,\Phi_3,\Phi_4),\nonumber\\
(\Phi_1,\Phi_3,\Phi_2,\Phi_4),\nonumber\\
(\Phi_1,\Phi_2,\Phi_4,\Phi_3),
\eea
we can simply apply, respectively, the following local
unitaries:
\bea
\frac{I-i\sigma_1}{\sqrt{2}} \otimes \frac{I-i\sigma_1}{\sqrt{2}},\nonumber\\
\frac{I+i\sigma_3}{\sqrt{2}} \otimes \frac{I-i\sigma_3}{\sqrt{2}},\nonumber\\
\frac{I+i\sigma_1}{\sqrt{2}} \otimes \frac{I-i\sigma_1}{\sqrt{2}}.
\label{eq:locper}
\eea

Therefore, any permutation $\sigma$ of the states (\ref{eq:max})
can be accomplished through local unitaries $T_i$, and any
Hamiltonian $H'$ satisfying conditions (\ref{eq:omaj})
[equivalently, conditions (\ref{eq:maj})] can be efficiently
simulated with $H$ and LU.

In the following we condense the previous findings into two
results, R1 and R2, which provide an explicit answer to problems
P1 and P2, respectively, announced in section II.C of the paper. We assume that the two--qubit
Hamiltonians $H$ and $H'$ are in their canonical form, with
$\vec{\lambda}$, $\vec{h}$, $\vec{\lambda'}$ and $\vec{h'}$ the
corresponding vectors of coefficients.

\vspace{2mm}

{\bf R1:} {\it Hamiltonian $H'$ can be efficiently simulated
by $H$ and LOCC ---or LU, LU+anc, or LO--- if and only if
conditions (\ref{eq:omaj}) [or, equivalently, conditions
(\ref{eq:maj})] are fulfilled, i.e. }
\be
H' \geq_{LOCC} H ~~\Leftrightarrow~~\vec{h'} \prec_s \vec{h}
~~\Leftrightarrow~~\vec{\lambda'} \prec \vec{\lambda}.
\ee

\vspace{2mm}

{\bf R2:} {\it The simulation factor $s_{H'|H}$ for LOCC ---or
LU, LU+anc, or LO--- protocols is given by the maximal $s>0$
such that $s\vec{h'} \prec_s \vec{h}$ or, equivalently, such
that $s\vec{\lambda'} \prec \vec{\lambda}$.}

\subsection{Explicit optimal LU protocols}

The last question we address is how to actually
construct a simulation protocol. That is, given $H$ and $H'$, we
show how to simulate $sH'$ using $H$ and LU, for
any $s\in [0,s_{H'|H}]$.

A complete answer to this question is given by a probability
distribution $\{p_k\}$ and a set of unitaries $\{u_k\otimes
v_k\}$ such that
\be
sH' = \sum_k p_k u_k\otimes v_k H u_k^{\dagger} \otimes
v_k^{\dagger},
\ee
where $s\in [0,s_{H'|H}]$, and $s_{H'|H}$ can be obtained using
result R2.

We already argued that it is always possible to choose all
$u_k\otimes v_k$ such that they permute the vectors of Eq.
(\ref{eq:max}), so that each $u_k\otimes v_k\equiv T_k$ is just
a composition of the local unitaries of Eqs. (\ref{eq:locper}).
As before, let $\{P_k\}_{k=1}^{24}$ denote the $24$ permutations
implemented by the local unitaries $\{T_k\}_{k=1}^{24}$. Then
the above problem reduces to finding an explicit probability
distribution $\{p_k\}$ such that
\be
s_{H'|H} H' = \sum_k p_k T_k H T_k^{\dagger},
\ee
or, equivalently, such that
\be
s_{H'|H} \vec{\lambda'} = \sum_k p_k P_k \vec{\lambda}.
\label{eq:decompo}
\ee
This is done on appendix B using standard techniques of convex
set theory. There we show how to construct a solution involving
at most 4 terms $p_kT_k$ for $s<s_{H'|H}$, and at most 3 terms
for optimal simulation, that is, when $s=s_{H'|H}$.

\section{Conclusions}

In this paper we have studied Hamiltonian simulation under the
broader scope of LOCC protocols. We have focused on
infinitesimal--time simulations, for which we have shown that
LOCC protocols are equivalent to LU+anc protocols, also that
LU+anc protocols are in general inequivalent to LU protocols (two--qubit Hamiltonians being an exception). For two--qubit Hamiltonians we have
rederived and extended the results of \cite{wir}, to finally
provide the optimal solution using LOCC.

Thus, the problem of simulating Hamiltonian evolutions has
received a complete answer for infinitesimal times and using
LOCC, for the simplest case of two-qubit systems. Several
interesting questions remain open. On the one hand,
the generalization of these results to systems other than two
qubits appears as challenging. On the other hand, the asymptotic
scenario for Hamiltonian simulation, where $H$ is used to
simulate $H'$ many times on different systems, certainly
deserves a lot of attention. 

Finally, we note that entangled ancillary systems have been recently shown to be of interest in non--local Hamiltonian simulation \cite{new}. In particular, entanglement can act a catalyst for simulations, both in the infinitesimal--time and finite--time regimes, in that in the presence of entanglement better time efficiencies can be obtained, although the entanglement is not used up during the simulation but is fully recovered after the manipulations.

\vspace{1cm}

The authors acknowledge discussions on the topic of Hamiltonian simulation with Charles H. Bennett, Debbie W. Leung, John A. Smolin and Barbara M. Terhal. Valuable and extensive comments from an anonymous referee are also acknowledged. This work was supported by the Austrian Science Foundation under
the SFB ``control and measurement of coherent quantum systems''
(Project 11), the European Community under the TMR network
ERB--FMRX--CT96--0087, the European Science Foundation, the
Institute for Quantum Information GmbH. and the National Science Foundation (of USA) through grant No. EIA-0086038. G.V also acknowledges a
Marie Curie Fellowship HPMF-CT-1999-00200 (European Community).
\appendix

\section{Extreme points of the set non--local Hamiltonian simulations achievable by LU+anc}

In this appendix we show that in LU+anc simulations
any Hamiltonian of the form
\be
H' = \bra{0_{A'} 0_{B'}} U\otimes V (H\otimes I_{A'B'})
U^{\dagger} \otimes V^{\dagger} \ket{0_{A'} 0_{B'}}
\label{eq:extH'}
\ee
can be efficiently simulated by $H$, for any couple of
unitaries $U$ and $V$ acting on $AA'$ and $BB'$. The result is
valid also for more than two systems after a straightforward
generalization of the following proof.

Notice that we can always write $U$ and $V$ using product basis
$\{\ket{i_{A}j_{A'}}\}$ and $\{\ket{i_{B}j_{B'}}\}$ as
\bea
U &=& \sum_{i=0}^{d_A-1} \sum_{j=0}^{d_{A'}-1}
\ket{i_{A}j_{A'}}\bra{\phi_{ij}}\\ V &=& \sum_{i=0}^{d_B-1}
\sum_{j=0}^{d_{B'}-1} \ket{i_{B}j_{B'}}\bra{\psi_{ij}},
\eea
where $\{\ket{\phi_{ij}}\}$ and $\{\ket{\psi_{ij}}\}$ are other
orthonormal basis of systems $AA'$ and $BB'$, respectively, and $d_{\kappa}$ denotes the dimension of system $\kappa$.

To perform this simulation, we need to make the output of the ancilla be the state $\ket{0_{A'} 0_{B'}}$, unentangled with the systems $AB$. This can not be achieved by performing just transformations $U$ and $V$, but by considering also a series of local unitaries $\{U_a\otimes
V_b\}$, $a\in\{0,\cdots, d_{A'}-1\}$, $b\in\{0,\cdots, d_{B'}-1\}$,
\bea
U_a \equiv I \otimes
(\sum_{l=0}^{d_{A'}-1}e^{i2\pi\frac{al}{d_{A'}}}
\proj{l_{A'}})U, \\ V_b \equiv I \otimes
(\sum_{l=0}^{d_{B'}-1}e^{i2\pi\frac{bl}{d_{B'}}}
\proj{l_{B'}})V,
\eea
and a constant probability distribution $\{p_{ab}\}$, $p_{ab}=
1/(d_{A'}d_{B'})$. Then we have that
$U_a^{\dagger}\ket{0_{A'}}=U^{\dagger}\ket{0_{A'}}$, and that
$\sum_a U_a = d_{A'} \proj{0_{A'}}U$, and similarly for $V_b$,
so that we obtain
\bea
\sum_{ab} p_{ab} U_a\otimes V_b (H\otimes I_{A'B'}) U_a^{\dagger}\otimes V_b^{\dagger}
\ket{0_{A'}0_{B'}} =\nonumber\\
\proj{0_{A'} 0_{B'}} U\otimes V (H\otimes I_{A'B'}) U^{\dagger}\otimes
V^{\dagger}\ket{0_{A'}0_{B'}}.\label{eq:lastpro}
\eea
Therefore Eq. (\ref{eq:lastpro}) defines a protocol that
simulates the Hamiltonian of Eq. (\ref{eq:extH'}) with unit time efficiency.

\section{Explicit two-qubit LU simulation protocols}

In this appendix we show how to find a probability distribution
$\{p_k\}$ and permutations $\{P_k\}$ such that
\be
\vec{\mu} = \sum_k p_k P_k \vec{\lambda},
\ee
for any two given four-dimensional, real vectors $\vec{\lambda}$
and $\vec{\mu}$ ($\vec{\mu} = s\vec{\lambda'}$ in section V.B)
such that $\vec{\mu} \prec \vec{\lambda}$, where $\sum_{i=1}^4
\lambda_i = \sum_{i=1}^4 \mu_i = 0$.

We first note two facts that will allow us to use standard
techniques of convex set theory: $(i)$ the set $S \equiv
\{\vec{\tau}~|~ \vec{\tau} \prec \vec{\lambda}\}$ is convex, and
($ii$) $\{P_k\vec{\lambda}\}_{i=1}^{24}$ are the extreme points
of $S$, as it follows from Birkhoff's theorem \cite{maj}. We can
then proceed as follows.

Step ($a$): we check whether $\vec{\mu} = P_i \vec{\lambda}$ for
any $i=1,\cdots,24$. If we find one such permutation we are
done. Otherwise we move to step ($b$).

Step ($b$): Facts $(i)$ and $(ii)$ guarantee that there is at
least one permutation $P_k$, that we call $Q_1$, and a positive
$\epsilon>0$ such that
\be
\vec{\mu} = \epsilon Q_1 \vec{\lambda} + (1-\epsilon)\vec{\tau},
\ee
where $\vec{\tau}$ also belongs to $S$, and therefore satisfies
$\vec{\tau} \prec \vec{\lambda}$. In other words, we have to
search until we find a permutation $Q_1$ such that
\be
(\vec{\mu} - \epsilon Q_1 \vec{\lambda})/(1-\epsilon) \prec
\vec{\lambda},
\label{eq:desig}
\ee
for some $\epsilon>0$. Once we have found it we only need to
increase $\epsilon$ to its maximal value compatible with Eq.
(\ref{eq:desig}). Let $q_1$ be this maximal value of $\epsilon$.
Then we can write
\be
\vec{\mu} = q_1 Q_1 \vec{\lambda} + (1-q_1)\vec{\mu}_2,
\ee
where $\vec{\mu}_2 \prec \vec{\lambda}$ is on one of the
surfaces of $S$ ---otherwise we could have taken a greater
$q_1$.

Such a surface is, again, a (lower dimensional) convex set,
whose extreme points are some of the $P_k\vec{\lambda}$'s, and
whose elements $\vec{\tau}$ fulfill $\vec{\tau}\prec
\vec{\lambda}$ but with one of the majorization inequalities
replaced with an equality. This allows us to repeat points $(a)$
and $(b)$, but now aiming decomposing $\vec{\mu}_2$ as a convex
sum of vectors $P_k\vec{\lambda}$. That is, first we check
whether $\vec{\mu}_2$ corresponds to $P_k\vec{\lambda}$ for some
$k$. And, if not, we search until we find a permutation $P_k$,
let us call it $Q_2$, such that, again,
\be
(\vec{\mu}_2 - \epsilon Q_2 \vec{\lambda})/(1-\epsilon) \prec
\vec{\lambda}.
\label{eq:desig2}
\ee
The maximum value of $\epsilon$ compatible with this equation,
say $q$, leads to a second term $q_2Q_2$ ($q_2=(1-q_1)q$) for
the decomposition of $\vec{\mu}$, 
\be
\vec{\mu} = (q_1Q_1+q_2Q_2)\vec{\lambda}+(1-q_1-q_2)\vec{\mu}_3,
\ee
and to a new $\vec{\mu}_3$, that lies
on a surface of yet lower dimensionality of the original
convex set $S$. We iterate the procedure until the remaining
vector $\vec{\mu}_l$ lies on a convex surface of $S$ of
dimension zero, which means that the surface contains only one
element, $\vec{\mu}_l$. In this way we obtain the desired
decomposition,
\be
\vec{\mu} = \sum_{k=1}^l q_kQ_k \vec{\lambda}.
\ee
What is the minimal value of $l$? For non-optimal simulation
protocols we have that $\vec{\mu} = s\vec{\lambda}'$, where $s<s_{H'|H}$, and
$\vec{\mu}$ is in the interior of $S$, which is a
three-dimensional set. Therefore the above procedure has to be
iterated at most three times before we are left with a
zero-dimensional surface of $S$, and the minimal decomposition
contains at most $l=4$ terms. For optimal simulation
protocols $\vec{\mu} = s_{H'|H}\vec{\lambda}'$ is already in a
surface of $S$, and therefore the minimal decomposition contains from $1$ to $3$ terms.

%\end{references}


\begin{thebibliography}{20}
\bibitem{note1} In this work we assume that local ancillary systems are available at wish in LO protocols (see section III), and thus we make no distinction between LO and LO+anc (as opposed to \cite{wir}).
\bibitem{neglect} As explained in \cite{wir}, local terms such as $I_{A}\otimes O_B$ can be added and removed by means of local unitary operations, and thus we can safely omit them in the present argumentation.
\bibitem{robustness} One can use the results of \cite{VJN}, concerning the robustness of entanglement transformations under a perturbation of the state under consideration, to show that the argument of section III.B.2 of this paper is also robust under $\O(t)$ modifications of a maximally entangled state.
\end{thebibliography}
\end{document}